\documentclass[]{raa}            

\usepackage{graphicx,times}             
\usepackage{natbib,amsmath}
\bibpunct{(}{)}{;}{a}{}{,}

\usepackage[a4paper=true,dvipdfm=true,pagebackref=true]{hyperref}

\usepackage{tabularx, booktabs, threeparttable}

\newcolumntype{L}[1]{>{\raggedright\arraybackslash}p{#1}}
\newcolumntype{R}[1]{>{\raggedleft\arraybackslash}p{#1}} 
\newcommand{\km}{${\rm km}\,{\rm s}^{-1}$}
\newcommand{\halpha}{H$_{\alpha}$}
\newcommand{\hi}{H\,{\sc i}}
\newcommand{\angs}{${\textrm{\AA}}$}

\def\apjl{Astrophysical Journal Letters}
\def\aap{Astronomy and Astrophysics}

\def\apjs{The Astrophysical Journal Supplement Series}

\begin{document}

\title{\textbf Interacting system NGC 7805/6 (Arp 112) and its tidal dwarf galaxy candidate
}


   \volnopage{Vol.0 (20xx) No.0, 000--000}      
   \setcounter{page}{1}          

   \author{ZhenXing Fu
      \inst{1,2}
   \and Chandreyee Sengupta
      \inst{1}
   \and Ramya Sethuram
      \inst{3}  
   \and Bikram Pradhan
       \inst{4}
   \and Mridweeka Singh
       \inst{4,5}
      \and Kuntal Misra
       \inst{4} 
    \and Tom. C. Scott
       \inst{6}       
    \and Yin-Zhe Ma
       \inst{7,8,1}
   }

   \institute{Purple Mountain Observatory (CAS), No. 8 Yuanhua Road, Qixia District, Nanjing 210034, China 
   {\it sengupta.chandreyee@pmo.ac.cn}\\
        \and
            School of Astronomy and Space Sciences,University of Science and Technology of China, Hefei 230026,China\\
        \and
            Indian Institute of Astrophysics, Koramangala, Bangalore 560034, India \\
        \and
             Aryabhatta Research Institute of Observational Sciences, Manora Peak, Nainital 263001, India\\
	\and
        Korea Astronomy and Space Science Institute, 776 Daedeokdae-ro, Yuseong-gu, Daejeon 34055, Republic of Korea \\
        \and 
            Institute of Astrophysics and Space Sciences (IA), Rua das Estrelas, 4150-762 Porto, Portugal\\
        \and
School of Chemistry and Physics, University of KwaZulu-Natal,
Westville Campus, Private Bag X54001, Durban, 4000, South Africa \\        
          \and  NAOC-UKZN Computational Astrophysics Centre (NUCAC), University of KwaZulu-Natal, Durban, 4000, South Africa\\
\vs\no
   {\small Received~~20xx month day; accepted~~20xx~~month day}}

\abstract{We present results from our Giant Metrewave Radio Telescope (GMRT) \hi\ , Himalayan Chandra Telescope (HCT) H$\alpha$, 1m Sampurnanand Telescope (ST) and 1.3m Devasthal Fast Optical Telescope (DFOT) deep optical observations of NGC 7805/6 (Arp 112) system to test KUG 2359+311's tidal dwarf galaxy (TDG) candidacy and explore the properties of the interacting system. Our GMRT \hi\ map shows no \hi\ detection associated with KUG 2359+311, nor any \hi\ tail or bridge-like structure  connecting KUG\,2359+311 to the NGC 7805/6 system.  Our HCT H$_{\alpha}$ image  on the other hand, shows strong detections in KUG 2359+311, with net SFR $\sim$  0.035$\pm 0.009 {\rm M}_{\odot}\,{\rm yr}^{-1}$. The H$\alpha$ data constrains the redshift of KUG 2359+311 to  $0.00 \le z \le 0.043$, compared to the redshift of NGC 7806 of $\sim 0.015$. TDGs detected to date have all been \hi\ rich, and displayed  \hi, ionised gas and stellar tidal debris trails (bridges or tails) linking them to their parent systems. But neither our \hi\ data nor our optical images, while three magnitudes deeper than SDSS, reveal tidal trail connecting KUG 2359+311 to NGC 7805/6. Lack of \hi\ , presence of an old stellar population,  ongoing star formation, reasonably high SFR  compared to normal dwarf galaxies suggest that KUG 2359+311 may not be an Arp\,112 TDG. It is most likely a case of a regular gas-rich dwarf galaxy undergoing a morphological transformation after having lost its entire gas content to an interaction with the Arp 112 system. Redshift and metallicity from future spectroscopic observations of KUG\,2359+311 would help clarify the nature of this enigmatic structure.}

   \authorrunning{Fu et al. }            
   \titlerunning{NGC 7805/6 and TDGs}  

   \maketitle

%


%
\section{Introduction}           

\noindent Tidal interactions between galaxies, where at least one of them is gas-rich, can result in tidal stripping of large amounts of \hi\  from the parent galaxy(s). This stripped gas then evolves and may form temporary structures which are likely to eventually fall back  into one or the
other of the interacting galaxies. If the gas mass and densities are sufficient and environmental conditions
are favourable, self-gravitating bodies with masses typical of dwarf galaxies, called Tidal Dwarf Galaxies (TDG), may form within the tidally stripped gas \citep[]{duc99,duc00}. To date, TDG candidates have been found almost exclusively within tidal debris of interacting galaxies. However, there remain significant unanswered questions about their independent existence as dwarf galaxies, their star formation (SF) properties and dark matter (DM) content. Atomic and molecular gas observations
of TDGs provide information for determining the conditions under which star formation is triggered within the tidally stripped gas, the role of gas in the formation of stars and the influence of dark matter in galaxy formation.

\noindent To address the issue of in--situ star and TDG formation in tidal debris, a multi-wavelength study of a sample of Arp interacting galaxies, Spirals, Bridges, and Tails (SB\&T), was carried out \citep[]{smith07,smith10}. GALEX, Spitzer and H$\alpha$ observations of SB\&T systems provided evidence of in--situ star formation and TDG candidates, including in the Arp\,112 interacting pair. Arp\,112 consisting of the galaxies NGC 7805 and NGC 7806 (radial velocities 4811 \km\ and 4768 \km\ respectively)   display signatures of a recent tidal interaction. Additionally, a smaller arc-like structure, KUG 2359+311,  is projected  $\sim$ 1 arcmin (20 kpc)  east of the principal pair (Figure 1). No spectroscopic data is available for  KUG\,2359+311 and thus it is not sure whether the structure is associated with Arp\,112 or not. If KUG 2359+311 is associated with Arp 112 system, then it could be a candidate TDG or remnant of a ring structure formed as a result of NGC\,7805/6 collision \citep{smith10}.  NGC\,7806 has an optical tail extending northward which contains two UV-bright clumps. These clumps are also potential TDG candidates or star-forming regions, but to date, neither clump has a published redshift \citep{smith10}.

\noindent We present here results from our Giant Metrewave Radio Telescope (GMRT) \hi\ line, Himalayan Chandra Telescope (HCT) H$\alpha$,  1m Sampurnanand Telescope (ST) and 1.3m Devasthal Fast Optical Telescope (DFOT) deep optical observations of Arp 112 system with an aim to test KUG 2359+311's TDG candidacy and explore the properties of the interacting system.
Using the average heliocentric velocity of the two principal galaxies from NASA Extragalactic Database
(NED), assuming $H_0=68\,{\rm km}\,{\rm s}^{-1}\,{\rm Mpc}^{-1}$~\citep{Planck-18}, we adopt a distance of 70
Mpc to Arp\,112. At this distance the spatial scale translates to $\sim 20.3\,{\rm kpc}\,{\rm arcmin}^{-1}$.

\section{Observations and Data Analysis}

\subsection{GMRT \hi\ line observations}
\noindent  The 21 cm \hi\ line observations of Arp 112 were carried out with the GMRT on February 1st, 2015. The
baseband bandwidth  for the observations was 16 MHz  resulting in a velocity resolution $\sim$7km/s.
Further details of the observations are given in Table 1.  The
Astronomical Image Processing System ({\sc aips}) software package was used to reduce the UV data. Bad data
from malfunctioning antennas and radio frequency interference (RFI) were flagged. The flux density calibration
scale used was \cite{baars77}, with flux density uncertainties $\sim$5\%. After calibration, the 
continuum subtraction was carried out using the {\sc aips} task {\sc uvlin}. The task {\sc imagr} was applied to
the visibilities to transform them into the final \hi\ image cubes. The integrated \hi\ , \hi\ velocity field and
velocity dispersion maps were made applying the task {\sc momnt} on the \hi\ cubes. Further details of the final map, presented in this paper, are given in Table \ref{table1}.

\begin{table}
\centering
\begin{minipage}{110mm}
\caption{GMRT observational and \hi\  map parameters}
\label{table1}
\begin{tabular}{ll}
\hline
Rest frequency & 1420.4057 MHz \\
Observation Date &1st February, 2015\\
Integration time  & \textcolor{black}{12.0 hrs } \\
Primary beam & 24\arcmin ~at 1420.4057 MHz \\
Map resolution (beam--FWHP)  &36.6$^{\prime\prime}$ $\times$ 36.0$^{\prime\prime}$, PA = -67.9$^{\circ}$ \\
Map rms per channel  & 1.0 mJy beam$^{-1}$  \\
RA (pointing centre)&\textcolor{black}{ 00$^{\rm h}$ 01$^{\rm m}$ 28.4$^{\rm s}$  }\\
DEC (pointing centre)& \textcolor{black}{ 31$^\circ$ 26$^\prime$ 16.0$^{\prime\prime}$}\\

\hline
\end{tabular}
\end{minipage}
\end{table}

\begin{figure}
\begin{center}
\includegraphics[width=15.8cm]{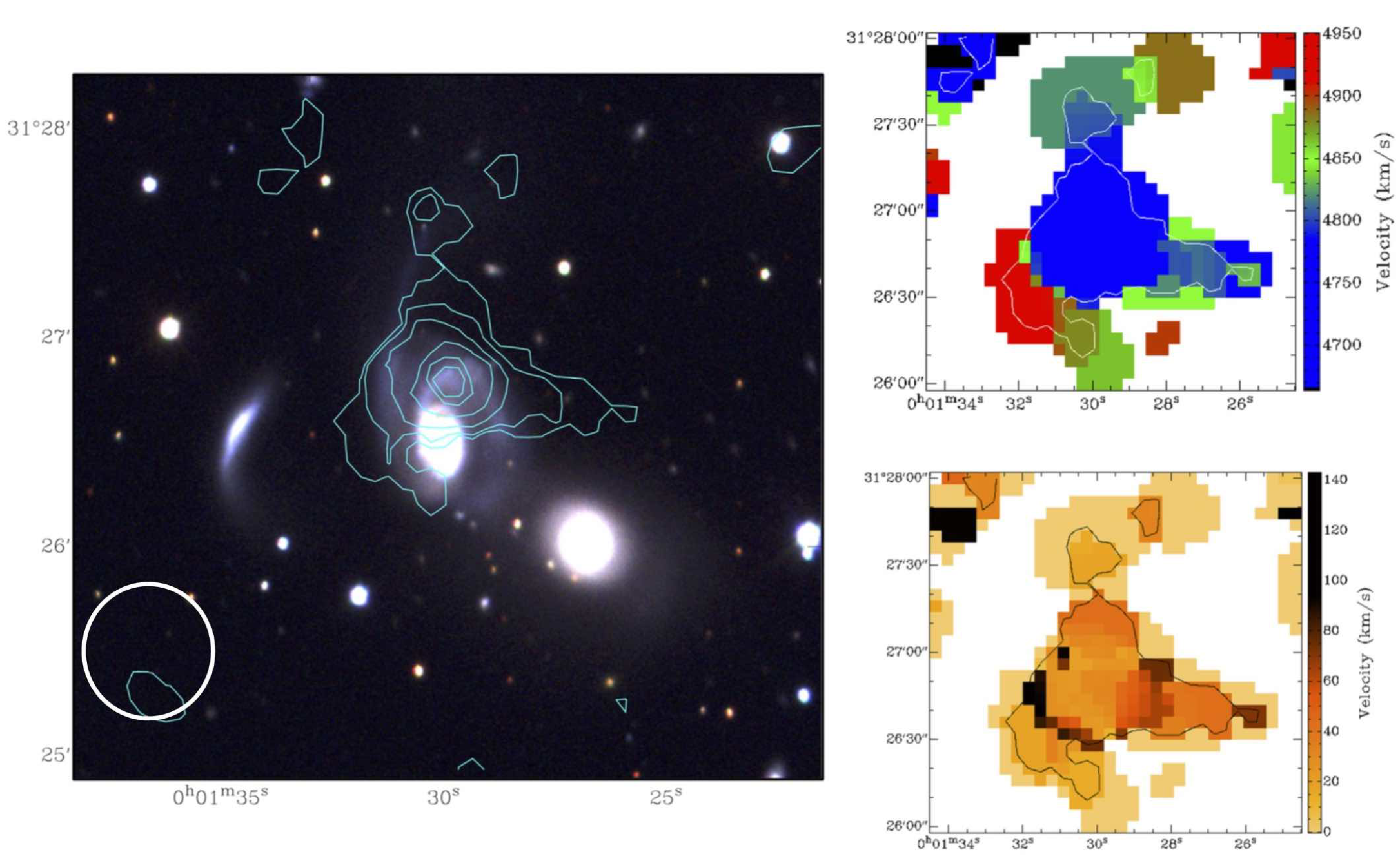} 
	\caption{{\bf Left panel:} Interacting pair NGC 7805/6 (Arp 112) field is shown here (x axis is RA and y axis is declination). Low resolution ($\sim$36.6$^{\prime\prime}\times$ 36.0$^{\prime\prime}$, PA -67.9$^{\circ}$) integrated \textcolor{black}{GMRT HI map contours overlaid on SDSS g, r, i composite}  image.  The beam is shown in a white ellipse in the lower left corner.  HI is detected only in NGC 7806. No HI is detected either in NGC 7805 or in KUG 2359+311, the cresent shaped structure to the east of the Arp 112 system. The HI column density levels are N(HI)=10$^{20}$ cm$^{-2}$ $\times$ (0.5,0.7, 1.2, 2.1, 3.0, 3.4). {\bf Upper Right panel:} The HI velocity field of NGC 7806. The white contour represents the lowest contour of the integrated HI map from left panel. This contour outlines the extent of the detected HI in NGC 7806. {\bf Lower Right panel:} The HI velocity dispersion map is shown here. The area shown here is similar to the HI velocity field and the black contour represents the extent of the detected HI in the galaxy. }
\label{fig1}
\end{center}
\end{figure}

\subsection{HCT H$_{\alpha}$ and optical broadband observations}

\noindent \halpha +[N{\sc ii}] observations of Arp\,112 system were obtained from 2m--Himalayan Chandra Telescope (HCT), installed at Indian Astronomical Observatory (IAO), Hanle and remotely controlled at CREST, Indian Institute of Astrophysics, Bengaluru. The data were obtained on  July 4th, 2016 using the
Himalaya Faint Object Spectrograph Camera (HFOSC), equipped with a 2K $\times$ 4K SITe CCD chip, but only the central 2K $\times$ 2K region (field of view of 10 $\times$ 10 arcmin$^2$) was used for imaging. The plate scale is 0.296 arcsec pixel$^{-1}$. Images were obtained using a broad H$_{\alpha}$ filter with a bandwidth of 500\angs. At the redshift of Arp\,112, the emission lines  [N II] 6549.9\angs, [N II] 6585.3\angs\ and H${\alpha}$ 6564.6\angs\  all fell within the bandwidth of \textcolor{black}{this} filter. $R$ band observations were also obtained \textcolor{black}{and used to subtract the continuum from the \halpha\ images.}  Several calibration frames, like the bias frames were obtained \textcolor{black}{throughout} the night and flat field frames for correcting pixel-to-pixel variation of CCD, were taken during morning and evening twilight. A spectrophotometric standard \textcolor{black}{star}, Feige 110, was also observed \textcolor{black}{on} the same night \textcolor{black}{to flux calibrate the images.}

\noindent The data obtained from HCT were processed \textcolor{black}{using}  {\sc iraf}\footnote{Image Reduction and Analysis Facility (IRAF) is distributed by the National Optical Astronomy Observatory, which is operated by the Association of Universities for Research in Astronomy (AURA) under a cooperative agreement with the National Science Foundation.} \textcolor{black}{software} following the standard procedures. The standard steps involved in data reduction are bias-subtraction, flat-field correction and alignment of the frames using {\sc iraf} tasks {\sc geomap} and {\sc geotran}. The point spread function (PSF) of the H${\alpha}$+[N{\sc ii}] image was matched to that of the R-band image using the {\sc gauss} task. The full width at half-maximum of the PSF, estimated from the stars in the $R$--band image, is $\sim$6 pixels (1.8 arcsec). As noted by \cite{james2004}, scaled R-band exposures give excellent continuum subtraction for observations made during nights. The multiple exposures were then summed to achieve a better signal-to-noise ratio (SNR) for \textcolor{black}{the} $R$ and H${\alpha}$ frames. The total on-target integration \textcolor{black}{time} achieved in $R$--band  \textcolor{black}{was}  600s and in the H${\alpha}$ band  3000s. The sky background in individual target frames was estimated from regions away from the galaxies and not affected by stars, which were then subtracted as a constant count from the frames. The frames were also divided by the total integration time to obtain per second images. The H$_{\alpha}$+[N{\sc ii}] line image was obtained after subtracting the PSF matched and scaled $R$--band image, \textcolor{black}{which also includes some line emission,} from the H${\alpha}$+[N{\sc ii}] image as described in \cite{scottarp202} and references therein. Using Feige 110 \citep{oke1990}, the flux conversion factor was estimated to be $5.47\times 10^{-16}\,{\rm erg}\,{\rm s}^{-1}\,{\rm cm}^{-2}$ / (count sec$^{-1}$) similar to the values obtained by \cite{scottarp202}, \cite{ramya2007,ramya2009}.


\subsection{ARIES ST and DFOT deep optical observations}

\noindent Observations of Arp\,112 were initiated in October, 2014 and continued \textcolor{black}{through} to January, 2015 \textcolor{black}{using the} 1m Sampurnanand Telescope (ST) and the 1.3m Devasthal Fast Optical Telescope (DFOT) of the Aryabhatta Research Institute of Observational Sciences (ARIES), India.  \textcolor{black}{Observations of Arp\,112 were made with these telescopes using} B, V, R and I broadband filters. Pre-processing of the images acquired with these two telescopes were \textcolor{black}{carried out} using standard tasks in {\sc iraf}.  Images were aligned in respective bands to achieve a high SNR.  We used the python module  {\sc alipy}  to align the images and the {\sc iraf}  task {\sc imcombine} \textcolor{black}{to} combine them.  The effective exposure time in each band is given in Table \ref{table2}.

\begin{figure}
\begin{center}
\includegraphics[width=14.8cm]{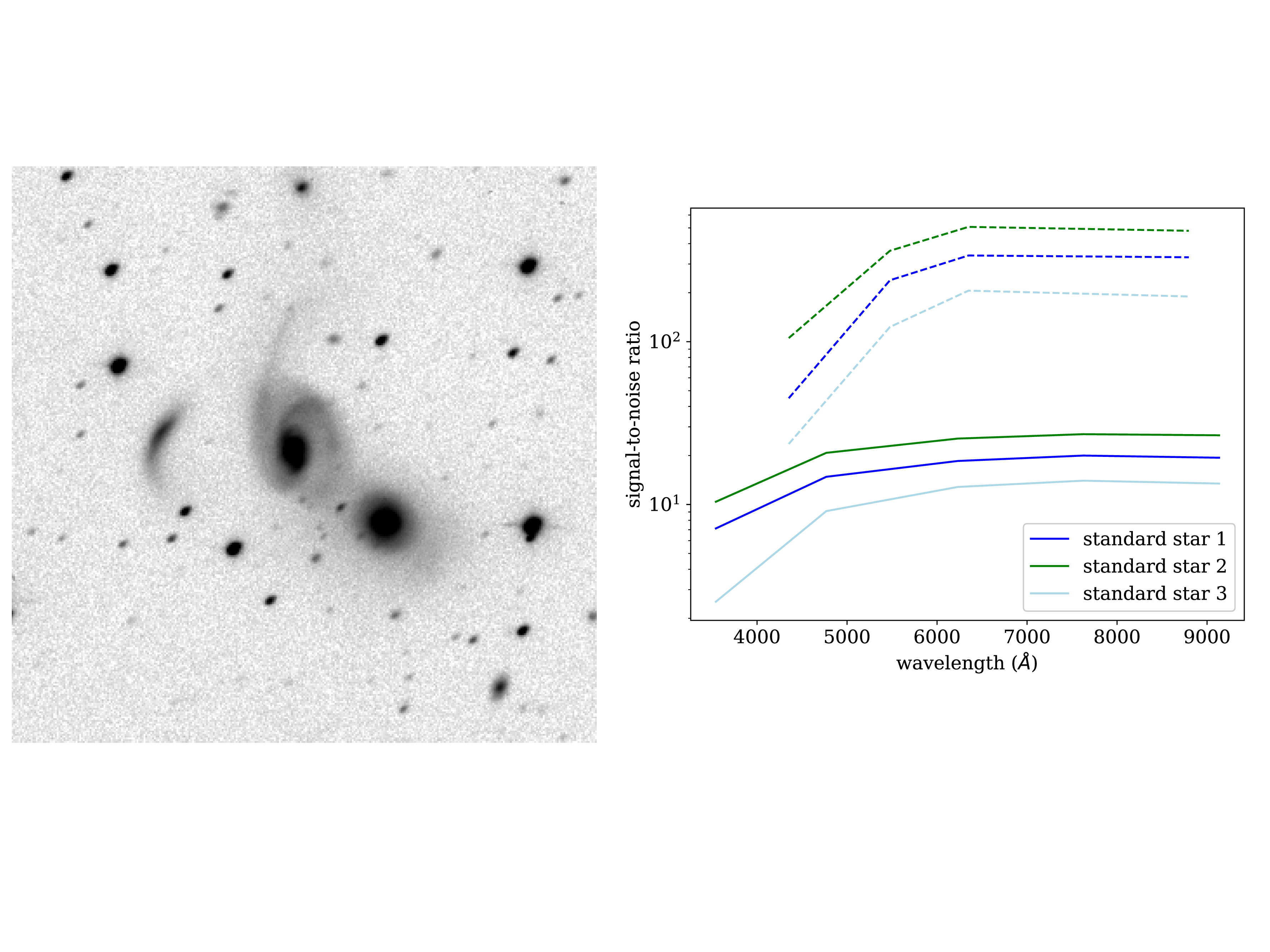} 
	\caption{{\bf Left}: Optical R-band image of the Arp 112 field, obtained by combining data from the 1m ST and the 1.3m DFOT. Field of view presented here is similar to the SDSS field in Figure 1.  {\bf Right:} SNR comparison of the 1m ST observation with the SDSS images. The solid lines and dashed lines correspond to SDSS and 1m telescope images, respectively. The ST images show  a better SNR in all the wavelength ranges. For longer wavelengths ($\ge$6500$\AA$) the images obtained from using the ST are $\sim$3 magnitude deeper than their SDSS counterparts.}
\label{fig2}
 \end{center}
\end{figure}

\noindent A photometrically non-variable source was required to compare observations at different epochs. None of the existing \textcolor{black}{standard star catalogues had an entry within a 30 arcmin radius of the Arp\,112}. Hence, the SDSS images from DR12 \citep{alam} were used to find the standard stars. Aperture photometry was performed over all sources present in the SDSS images in all five filters (u, g, r, i, z). The source flux F obtained from aperture photometry was converted to instrumental magnitudes using the expression, $m_{\rm inst}= -2.5\log_{10}(F/t)$, where t is the exposure time in SDSS frame $\sim$54 seconds. The offset between each instrumental magnitude and it’s respective SDSS magnitude was compared to find any sources that behaved inconsistently in each band.  1$\sigma$ clipping was done iteratively to remove the outlier sources, and three sources were selected randomly to compare our observations with those from SDSS. These three sources were considered in our observation as well as SDSS frames to derive SNRs. The calculated  SNR for the three sources is shown in Figure \ref{fig2}, with our observations having a better SNR in all the observed wavelength bands compared to SDSS. For longer wavelengths ($\ge$6500\angs) the images obtained using  \textcolor{black}{the Aries telescopes were} $\sim$3 magnitudes deeper than the SDSS images.


\begin{table}
\centering
\begin{minipage}{110mm}
\caption{ST and DFOT observations: effective exposure time per filter }
\label{table2}
\begin{tabular}{llll}
\hline
	Filter & Central wavelength ($\angs$) &Exposures (sec) & Effective exposure time (sec.)  \\
B	 & 4361 & 300 $\times$ 23&6900\\
V       &5448 & 300 $\times$ 73&21900\\
R	& 6407 & 300 $\times$ 4&1200 \\
I	& 7980 & 300 $\times$ 36& 10800\\
\hline
\end{tabular}
\end{minipage}
\end{table}


\begin{figure}
\begin{center}
\includegraphics[width=12.0cm]{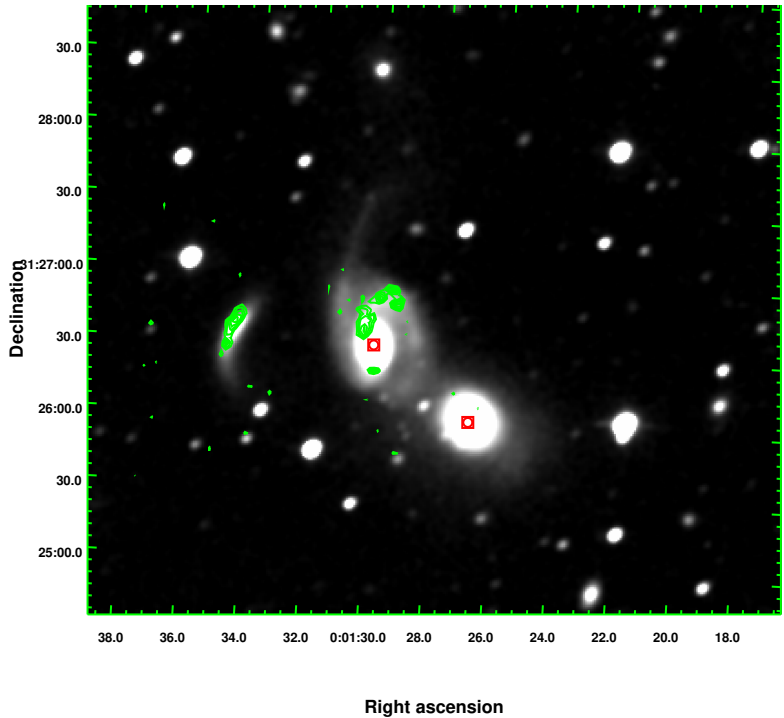} 
	\caption{Arp 112 HCT  H$\alpha$+[N II] image contours (green) overlaid on a deep optical image [R-BAND] from the ST  1m and DFOT 1.3m telescopes. Red squares mark the centres of NGC\,7805/6.  The contour levels are marked at 0.5$\sigma$, 1$\sigma$ and 1.5$\sigma$ above the background. }
\label{fig1}
\end{center}
\end{figure}

\begin{figure}
\begin{center}
\includegraphics[width=12.0cm]{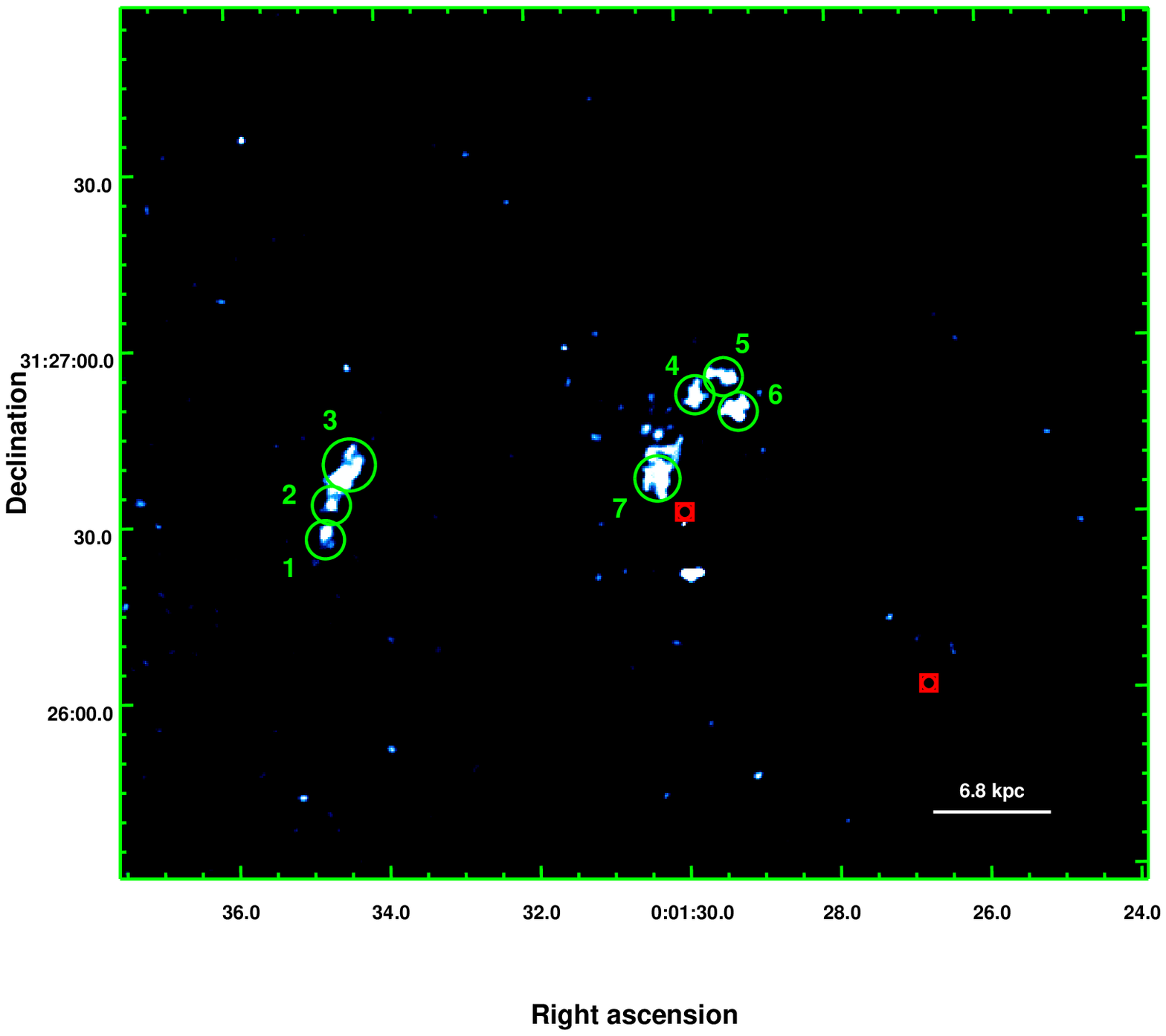} 
	\caption{Arp\,112 HCT  H$\alpha$+[N II] image showing 7 star forming  knots  identified in KUG 2359+311 (knots 1,2 and 3)  and NGC\,7806 (knots 4, 5, 6 and 7). The size of the apertures drawn around the emission varies from 2 kpc for knots 1,2,4,5,6,7 to 3 kpc for knot 3. The total size of the H$\alpha$ disk of KUG\, 2359+311 estimated by the extent of  all the knots is $\sim$7.5 kpc }
\label{fig1}
\end{center}
\end{figure}

\section{Results and Discussion}

\subsection{NGC 7085/6 system}

\noindent Figure 1 shows the \hi\ column density contours from the GMRT  (36.6$^{\prime\prime}\times$ 36.0$^{\prime\prime}$, PA -67.9$^{\circ}$) integrated \hi\ map for the Arp\,112 field overlaid on an SDSS optical image. The entire
 \textcolor{black}{\hi\ emission detected in the system} is associated with NGC\,7806. The deep optical image of NGC\,7806 \textcolor{black}{in Figure \ref{fig2} shows a faint tidal tail which extends to the north beyond the optical disk. Additionally, the northern disk of NGC 7806 hosts UV bright clumps identified in GALEX images \citep{smith10}. The \hi\ column density maxima are projected on this UV bright section of the  NGC\,7806  northern disk (Figure 1) and offset by $\sim$ 25$^{\prime\prime}$ (8 kpc) from the optical centre}. NGC\,7805, on the other hand, is devoid of \hi\ .  This \hi\ displacement in NGC 7806 is common in strongly interacting pairs.   Even larger displacements of the bulk of a pair's \hi\ has been observed other Arp interacting pairs, e.g. Arp 181 and Arp 65 \citep{seng13, seng15}. Arp\,112 has been observed \textcolor{black}{using the Arecibo 305m single-dish telescope (FWHM beam $\sim~3.5\,{\rm arcmin}$) and an \hi\ flux density of 
1.25 Jy \km\ was reported in \cite{jack}}. At the distance of the Arp 112, this translates to an \hi\ mass of $1.4 \times 10^{9}\,{\rm M}_{\odot}$. Using GMRT spectrum, we measured the \hi\ flux density to be 1.0 Jy \km\ and the corresponding \hi\ mass to be $1.1 \times 10^{9}\,{\rm M}_{\odot}$.  Given that uncertainties on these estimates are $\sim$ 10\%, this indicated that almost all of the \hi\ flux was recovered by the GMRT interferometric observations.

\noindent Since NGC\,7805 is an S0 type galaxy and NGC\,7806 is an Sbc spiral, and all the detected \hi\ is
associated to NGC 7806, it can \textcolor{black}{ reasonably be assumed that NGC\,7806 contained most of the pair's \hi\  before they started interacting. The velocity of the detected \hi\ (4751 \km) is closer to the optical radial velocity of NGC\,7806 (4768 \km) compared}  to NGC 7805 (4811 \km), \textcolor{black}{confirming}  the bulk of the detected \hi\  is most likely to have originated in  NGC\,7806. We attempt here to
\textcolor{black}{determine whether  NGC\,7806 shows signs of \hi\  loss due to the } interaction. A field galaxy of the same size and
morphology as NGC\,7806, \textcolor{black}{is expected to}  have an \hi\ content of $3.6 \times 10^{9}\,{\rm M}_{\odot}$~\citep{hayn84}. However, the measured total \hi\ mass for NGC\, 7806 is only $1.4 \times 10^{9}\,{\rm M}_{\odot}$  giving an \hi\ deficiency of 0.41. \textcolor{black}{  This \hi\ deficiency value suggests, that the NGC\,7805/06 interaction has removed a significant quantity of the NGC\,7806 \hi\ content. But it should be noted that the uncertainties for the expected \hi\ content of a single galaxy are high  \citep{hayn84}. Therefore we must be cautious with our conclusion about \hi\ deficiency  of NGC\,7806 and can only comment that it is likely that a significant \hi\ loss has}  occurred during the interaction.

\noindent \textcolor{black}{Figure 1 -- right panel, shows the \hi\ velocity field  with velocities per the  colour bar to the right.}  The black contour represents the lowest \hi\ column density contour of the  total intensity map. The black contour has been used to show the outline of the velocity field for the \hi\ total intensity map. The \textcolor{black}{\hi\ velocity field does not show any systematic velocity pattern, i.e.  no sign of any systematic rotation.} There are signs of irregular velocity gradients in the \hi\  but since the \hi\ velocity field is intensity weighted, the velocity gradient towards the low column density regions is not significant enough to be trusted.  The unmasked velocity field map further demonstrates the poor signal to noise ratio which prohibits us from drawing any conclusion about NGC 7806's kinematics.  \textcolor{black}{ The interaction with NGC 7805 appears to have  destroyed the regular kinematics of the  NGC 7806's \hi\ disk and the remnant \hi\  is now detected with its column density maxima at $\sim$ 8 kpc  north of the NGC\,7806 optical centre.}


 
\noindent Figure 3 shows the H$\alpha$ detection on the NGC 7805/6 system as well as in KUG 2359+311. NGC\,7806 and NGC\,7805 optical centres are marked with red squares and the H$\alpha$ detections are shown with green contours. NGC 7806, which is an Sbc galaxy shows at least 4 knots of star formation numbered from 4 to 7. No H${\alpha}$ detection is seen from NGC 7805, an S0  galaxy. Following \cite{ramya2007,ramya2009}, we selected the emission boundary based on the criteria that the emission is centrally peaked and the boundary of the region is set where the flux falls to 1$\sigma$ of the background (i.e. $\sim 10^{-17}\,{\rm erg}\,{\rm s}^{-1}\,{\rm cm}^{-2}$). H${\alpha}$+[N{\sc ii}] fluxes, luminosities and star formation rates (SFR) for each of these knots are estimated and tabulated in Table 3. The fluxes are in the range $10^{-15}$-$10^{-16}\,{\rm erg}\,{\rm cm}^2\,{\rm s}^{-1}$ and luminosities are in the range $10^{38-39}\,{\rm erg}\,{\rm s}^{-1}$ as shown in the Table 3. The SFR values are based on the formula from \cite{kennicutt1998} for galaxies with solar abundances and assuming Salpeter initial mass function with stellar masses ranging from $0.1$-$100\,{\rm M}_{\odot}$. For individual star-forming knots in NGC 7806, SFRs are in the range  $0.01\pm 0.005$ - $0.025 \pm 0.007\,{\rm M}_{\odot}\,{\rm yr}^{-1}$. The positions of these star-forming H$\alpha$ clumps match well the UV bright zones in the northern disk of NGC 7806, reported in \cite{smith10}. Interestingly  the \hi\ mass observed in NGC 7806 also has its column density peak projected at the position of these clumps. This suggests that this star formation could be triggered by the interaction of NGC 7806 and NGC 7805.

\begin{table}[h!]
  \begin{center}
    \caption{H$\alpha$ properties of the star forming knots of Arp 112 system.}
    \label{table3}
    \begin{tabular}{lllll}
      \hline
      \\
      {\bf{Knots}} &
      {\bf{Diameter}} &
      {\bf{F(H$\alpha$)}} &
      {\bf{L(H$\alpha$)}} &
      {\bf{SFR(H$\alpha$)}} 
      \\
         &
      kpc &
      $10^{-15}\,{\rm erg}\,{\rm s}^{-1}\,{\rm cm}^{-2}$ &
      $10^{39}\,{\rm erg}\,{\rm s}^{-1}$ &
      ${\rm M}_{\odot}\,{\rm yr}^{-1}$ \\
      
      \hline
      1 & 2.2 & 0.87$\pm$0.67 & 0.59$\pm$0.39 & $0.004\pm0.003$                  \\
      2 & 2.2 & 2.61$\pm$1.11 & 1.53$\pm$0.65 & $0.012\pm0.005$                   \\
      3 & 3.0 & 4.18$\pm$1.41 & 2.45$\pm$0.83 & $0.019\pm0.006$                  \\ 
	    &&&&  SFR (KUG 2359+311) \\
	    &&&&     $0.035\pm0.009$ \\
      4 & 2.2 & 2.46$\pm$1.08 & 1.44$\pm$0.64 & $0.011\pm0.005$  \\
      5 & 2.2 & 2.85$\pm$1.16 & 1.67$\pm$0.68 & $0.013\pm0.005$                 \\
      6 & 2.2 & 2.62$\pm$1.12 & 1.54$\pm$0.66 & $0.012\pm0.005$               \\
      7 & 2.6 & 5.19$\pm$1.58 & 3.05$\pm$0.93 & $0.024\pm0.007$               \\
	    &&&&  SFR (NGC 7806)   \\
	    &&&&$0.061\pm0.012$\\
      \hline
     \end{tabular}
  \end{center}
\end{table}


\subsection{KUG 2359+311, a TDG candidate?}

\noindent KUG 2359+311 is projected \textcolor{black}{$\sim$ 1 arcmin (20 kpc) east of the  NGC\,7805/6 pair and has been proposed} as a TDG candidate for Arp
112 \citep{smith10}. \textcolor{black}{To date no spectroscopic redshift  has been published for KUG 2359+311
 and thus it is not confirmed that the structure is associated with  Arp\,112}. While KUG\,2359+311 is clearly visible in the GALEX image \citep{smith10}, it also stood out as one of the \textcolor{black}{redest
 TDG canidates}  in the \cite{smith10} sample. The system seemed interesting for an \hi\ imaging study because \textcolor{black}{TDGs  in systems involving an early type parent galaxy are rare. Usually most
TDGs are found in pairs of interacting gas-rich  large spirals.}

\noindent Our GMRT \hi\ map shows \textcolor{black}{ no \hi\ detection associated with KUG 2359+311, nor } any \hi\ tail or bridge-like structure was detected connecting KUG\,2359+311 to the NGC 7805/6 system.  A search across the usable bandwidth of this observation, covering an approximate redshift range of  0.0116$\le$ z $\le$0.0214, yeilded no reliable \hi\ signal on or around the position of KUG 2359+311. Assuming it to be at the same redshift as Arp 112, a 5$\sigma$ upper limit to its \hi\ mass would approximately be 1.2$\times$10$^{8}$ $\rm M_{\odot}$. Our HCT H$_{\alpha}$ image \textcolor{black}{(Figures 3 and 4)} on the other hand, show strong detections in KUG 2359+311. Unfortunately, this H$\alpha$ detection \textcolor{black}{provides only a broad constraint on } KUG\,2359+311's redshift as the  \textcolor{black}{  \halpha\ filter used covers a relatively large  redshift range between $z=0.0$ and $0.043$, compared to the redshift of NGC 7806 of $\sim$ 0.015. So while we have been able to put a constraint on its redshift, whether KUG 2359+311 is associated with Arp\,112,  will require spectroscopic redshift confirmation. In the absence of unambiguous evidence that  KUG 2359+311 is  a TDG  associated with  Arp\,112, we studied the properties of KUG\,2359+311 from the available optical data and compared it with known dwarf galaxy properties.}

\noindent The H$\alpha$+[N{\sc ii}] emission from KUG 2359+311  shows three knots of star formation numbered as 1, 2 and 3 in Figure 4. The SFRs estimated for individual SF regions of KUG 2359+311 from the H$\alpha$ clumps are between $0.004\pm 0.003$ to $0.02\pm0.006\,{\rm M}_{\odot}\,{\rm yr}^{-1}$. The total SFR for KUG 2359+311 is estimated to be $0.035 \pm 0.009\,{\rm M}_{\odot}\,{\rm yr}^{-1}$. Following \cite{kennicutt1998}, the SFR derived from FUV magnitude for KUG 2359+311 system \citep{smith10}  is $\sim$ $0.042\pm 0.013\,{\rm M}_{\odot}\,{\rm yr}^{-1}$  matches well within the range of errors to the SFR estimated from our H$\alpha$ observations. This SFR range is normal for a star-forming dwarf galaxy \citep{jessica}.  Assuming KUG 2359+311 to be at the distance of Arp 112, the estimated size of the KUG 2359+311 optical and  H$_{\alpha}$ extent would be $\sim$ 12 kpc and 7.5 kpc respectively. Using SDSS colours and MLCR relation from \cite{2015MNRAS.452.3209R}, the estimated stellar mass for KUG 2359+311 at the distance of Arp 112 is $1.8\times 10^{9}\,{\rm M}_{\odot}$.  Under these assumptions, if KUG 2359+311 were to be at the same redshift as Arp 112, its optical mass, H$\alpha$ disk extent and SFR of  KUG 2359+311 are characteristic of a star-forming gas deficient dwarf galaxy rather than a recently formed TDG. Also, no sign of any diffuse tidal or stellar remnants or bridges were found in the deep optical image from the ARIES telescopes. 

\noindent \textcolor{black}{TDGs detected \textcolor{black}{to} date have all been \hi\ rich, and display  \hi, ionised gas and/or stellar tidal debris trails (bridges or tails) linking them to their  parent systems  \citep[]{duc99,duc00,seng13,seng14,seng17,smith07,smith10}. Moreover, their blue optical colours \citep[B-V =0.3][]{duc99} are consistent within the bulk of their stars having been formed within the last  $\sim$ 0.5 Gyrs \citep{scottarp202}.} Although proposed as a TDG candidate by \cite{smith10}, the lack of evidence linking  KUG 2359+311 to the NGC 7805/6 system and its stellar mass, presence of old stellar population,  size and SFR (assuming same redshift as Arp\,112) raise doubts about whether it is a TDG or not.  Until a spectroscopic redshift for KUG\,2359+311 is available the possibility that it is a foreground or background galaxy remains open. A further possibility is that it is an ordinary dwarf galaxy caught up in the interaction between the Arp\,112 pair, which could explain its unusual crescent morphology. If KUG 2359+311 was a dwarf or small irregular galaxy at the same redshift as Arp 112,  its ongoing star formation rate and its complete lack of \hi\ make it an interesting case of a dwarf undergoing a possible morphological transformation. The SFR of KUG 2359+311 estimated from GALEX  FUV  magnitude \citep{smith10}  is $\sim 0.042 \pm 0.013\,{\rm M}_{\odot}\,{\rm yr}^{-1}$ and that from the H$\alpha$ image is $0.035\pm0.009\,{\rm M}_{\odot}\,{\rm yr}^{-1}$. Within errorbars, they are same, though the absolute numbers suggest a decrease in SFR over time. The error bars on these estimates prevent us from making any strong claim. Still, a possible scenario could be that KUG 2359+311 was a normal dwarf which got caught in the Arp 112 interaction and went through an interaction induced star formation phase with the gradual depletion of \hi\ as well as its SFR. In its current state, it appears to have lost all its \hi\ content to either the IGM or to star formation. It is possible that we are witnessing a dwarf galaxy in a state of morphological transformation and that post this star-forming stage the galaxy may transform into a gas deficient, red early-type dwarf. This scenario seems more plausible as opposed to KUG 2359+311 being a foreground or a background galaxy due to the lack of any larger mass nearby galaxy that can explain its crescent-shaped distorted morphology. The redshift and metallicity from future spectroscopic observations of KUG\,2359+311 would help clarify the nature of this enigmatic structure.


\section{Conclusion}

\noindent We present our GMRT \hi\ , HCT H$\alpha$ ,  1m ST and 1.3m DFOT deep optical observations of an interacting system, Arp\, 112,
involving an S0 galaxy NGC 7805 and an Sbc spiral NGC 7806. Arp\,112 was \textcolor{black}{proposed to also}  contain
a TDG candidate KUG 2359+311, although, no spectroscopic data were available to support this. Our observations reveal that the \hi\ in the pair is now projected to the north of NGC\,7806 disk with the \hi\ column density maxima located about eight kpc north of the NGC\,7806  optical centre near an ongoing star formation region. Assuming that NGC\, 7806, was the source of the bulk of the detected \hi,\ we
conclude  NGC 7806 has a \hi\ deficiency of 0.41 suggesting it has lost a substantial fraction of its original \hi\ due to the interaction.  No \hi\ was detected in 
NGC 7805 or KUG 2359+311, the TDG candidate. Our optical imaging, while three magnitudes deeper than SDSS, did not reveal any additional tidal debris. On the other hand, our  H$\alpha$ observations revealed starforming regions coincident with the highest column density \hi\ in NGC\,7806 and along the length of  KUG 2359+311 optical structure.
The  H$\alpha$ filter used for the HCT observations provide only a broad constraint ($z=0.0$-$0.043$) on the redshift of KUG 2359+311. As a  result of this and our analysis of the optical data for KUG\,2359+311  we consider it is  prudent to remain open to the possibility that KUG 2359+311 is not an Arp\,112 TDG, but is most likely a case of a normal gas-rich dwarf galaxy undergoing a morphological transformation after having lost its entire gas content to an interaction with the Arp 112 system.

\section{Acknowledgements}
\noindent We thank the staff of the GMRT that made these observations possible. GMRT is run by the National Centre for Radio Astrophysics of the Tata Institute of Fundamental Research. We thank the staff of IAO, Hanle and CREST, Hosakote, that made these observations possible. The facilities at IAO and CREST are operated by the Indian Institute of Astrophysics, Bangalore. TS   acknowledges support by Funda\c{c}\~{a}o para a Ci\^{e}ncia e a Tecnologia (FCT) through national funds (UID/FIS/04434/2013), FCT/MCTES through national funds (PIDDAC) by this grant UID/FIS/04434/2019 and by FEDER through COMPETE2020 (POCI-01-0145-FEDER-007672).  TS  also acknowledges support from DL 57/2016/CP1364/CT0009.This research has made use of the NASA/IPAC Extragalactic Database (NED) which is operated by the Jet Propulsion Laboratory, California Institute of Technology, under contract with the National Aeronautics and Space Administration.This research has made use of the Sloan Digital Sky Survey (SDSS). The SDSS Web Site is http://www.sdss.org/. Y.Z.M. acknowledges the support of NRF with grant no.105925, 109577, 120385, and 120378, and NSFC with grant no. 11828301.

\bibliographystyle{raa}
\bibliography{cig}
\end{document}